\documentclass[%
  notitlepage,
  12pt,
  nofootinbib,
 amsmath,amssymb,
 pre,
]{revtex4-2}

\usepackage{graphicx}
\usepackage{dcolumn}
\usepackage{bm}

\newcommand{\iu}{{i\mkern1mu}}
\newcommand*\diff{\mathop{}\!\mathrm{d}}
\DeclareMathOperator{\erf}{erf}
\DeclareMathOperator{\erfi}{erfi}
\begin{document}

\title{Diffusion approximations in population genetics and the rate of Muller's ratchet}

\author{Camila Br\"autigam}
\affiliation{%
Max Planck Institute for Mathematics in the Sciences, Leipzig, Germany
}
\author{Matteo Smerlak}%
 \email{smerlak@mis.mpg.de}
\affiliation{%
Max Planck Institute for Mathematics in the Sciences, Leipzig, Germany
}%

\date{\today}

\begin{abstract}
Diffusion theory is a central tool of modern population genetics, yielding simple expressions for fixation probabilities and other quantities that are not easily derived from the underlying Wright-Fisher model. Unfortunately, the textbook derivation of diffusion equations as scaling limits requires evolutionary parameters (selection coefficients, mutation rates) to scale like the inverse population size---a severe restriction that does not always reflect biological reality. Here we note that the Wright-Fisher model can be approximated by diffusion equations under more general conditions, including in regimes where selection and/or mutation are strong compared to genetic drift. As an illustration, we use a diffusion approximation of the Wright-Fisher model to improve estimates for the expected time to fixation of a strongly deleterious allele, i.e. the rate of Muller's ratchet.
\end{abstract}

\maketitle

\section{Introduction}

The Wright-Fisher (WF) model captures in simple mathematical form the combined effects of natural selection, mutation, and genetic drift, and therefore plays a foundational role in quantitative evolutionary theory and population genetics. Unfortunately, the main quantities of interest such as fixation probabilities cannot be computed in closed form within this model. An established approximation strategy relates the discrete WF model to a continuous diffusion process in allele frequency space obtained via a scaling limit in which the population size and timescale go to infinity in a correlated manner. This scaling transformation was first used by Fisher \cite{Fisher_1923}, Wright \cite{wright1931} and Kimura \citep{Kimura_1955}, and more rigorous expositions were given by Kolmogorov \cite{Kolmogoroff_1931}, Malécot \cite{malcot1945} and Feller \cite{Feller_2015}. This approach plays a central role in most textbooks expositions \cite{Ewens_2004, rice2004evolutionary, nagylaki2013introduction, crow2017introduction}, and it has been said that ``the standard diffusion approximation has permeated the field so
thoroughly that it shapes the way in which workers think about the
genetics of populations'' \citep{wakeley2005}.

A key limitation of the WF diffusion equation, however, is its narrow range of applicability: for the deterministic forces of selection and mutation to
survive the scaling limit, selection coefficients and
mutation rates must scale like the inverse effective population size $1/N$. This condition is neither biologically well motivated (selection coefficients and mutation rates
do not covary with population size) nor universally applicable (viral populations, for instance, can have large effective sizes $\mathcal{O}(10^{7})$ \cite{lumby2020large}). Perhaps because the scaling limit of the WF model is often presented as \emph{the} diffusion approximation of population genetics, diffusion methods are sometimes believed to be restricted to the weak selection-weak mutation regime of evolution: ``In order to use diffusion approximations it is necessary that the mutation rates be of no larger order of magnitude than $1/N$'' \cite{Ewens_1965}. Outside this regime, the common practice is to either assume neutral evolution and effectively set all selection coefficients and mutation rates to zero, as in Kimura's original work \citep{Kimura_1955}, or to use deterministic equations \citep{brger2000}, \emph{i.e.} to neglect genetic drift altogether.

Nevertheless, there are important cases where genetic drift is small compared to selection and mutation, and yet plays a key role in driving evolutionary change. An example is \emph{Muller's
ratchet} \citep{Muller_1932,felsenstein1974}, \emph{viz.} the irreversible accumulation of slightly deleterious mutations in finite asexual populations. This process has been invoked in the evolution of mitochondria \citep{LOEWE_2006}, RNA viruses \citep{Chao_1990}, Y chromosomes \citep{GORDO_2001}, ageing \citep{Govindaraju_2020}, cancer \citep{lopez2020interplay}, and sex \citep{HARTFIELD_2012}. Obtaining analytical expressions for the mean click time of the ratchet is therefore an important goal for theory.

At first sight, diffusion theory seems ideally suited for this
problem: Muller's ratchet is akin to a fixation problem for a two-type Wright-Fisher process, where one type represents unloaded alleles, and the other type comprises all other, deleterious mutations \citep{Stephan_1993,GORDO_2001,Waxman_2010}. However, recent works have reinforced the belief that diffusion theory is inapplicable under finite selection strengths \citep{Assaf_2011,Metzger_2013}. An emerging consensus is that diffusion theory just does not apply to Muller's ratchet, at least not in the slow click regime: ``The fact that the extinction of the fittest class is due to {[}\ldots{}{]} a rare, large fluctuation and not {[}\ldots{}{]} a typical fluctuation prohibits simple diffusive treatments of the ratchet'' \citep{Metzger_2013}.\footnote{Ironically, Eq.~(23)
of \citep{Metzger_2013} coincides with the large $N$
Laplace approximation of the mean click time for the WF diffusion equation, Eq.~{\eqref{WFdiff}} below.}

This paper has two objectives. First, we stress that diffusion methods in population genetics do \emph{not} in fact require $s,u,v\sim N^{-1}$ ($s$ selection coefficient, $u,v$ mutation rates) to correctly capture the interplay between mutation and selection in large populations; an alternative form of the WF diffusion equation, obtained as an interpolation of the WF process rather than as a scaling limit, provides an excellent approximation valid under the more general conditions $s,u,v\ll 1\ll N$. The fact that diffusion theory can be used to derive more general results than the usual classical Wright-Fisher diffusion equation has been noted in the past (see e.g. \citep{wakeley2005} and references therein), but has not been sufficiently appreciated in our view. Second, we use the alternative approximation by diffusion to derive a new analytical formula for the mean fixation time of a deleterious allele, a central quantity in the context of Muller's ratchet. We find that not only can diffusion theory be applied to this problem, it actually provides a better approximation than previous estimates in \citep{Assaf_2011,Metzger_2013} based on WKB expansions.

\section{Scaling limit vs. interpolation}
We focus on the haploid WF model for two types $A$ and $B$. For a population with effective size $N$, this is the Markov chain on the set $\{0,1/N, \cdots, (N-1)/N, 1\}$ defined by the binomial sampling formula
\begin{equation}
    NX_{n+1} \sim\textrm{Binom}(N,p(X_{n})).
  \label{WF}
\end{equation}
Here $X_n$ denotes the frequency of $A$ types at generation $n$, and $p(x)$ represents the probability that an offspring is of type $A$
given that the frequency of $A$ in the parental
population is $x$. The form of the function $p(x)$ depends on the specifics of
mutation-selection dynamics. For instance, if $A$ has $(1+s)$ times more offspring than $B$, and the probability of a mutation $A\to B$ (resp. $B\to A$) is $u$ (resp. $v$), we have
\begin{equation}
   p(x) = \frac{(1+s)(1-u)x + v(1-x)}{1+sx}.
   \label{probability}
\end{equation}
Other functional forms are more appropriate for viability selection, etc.

\begin{figure}
    \includegraphics[width=.9\columnwidth]{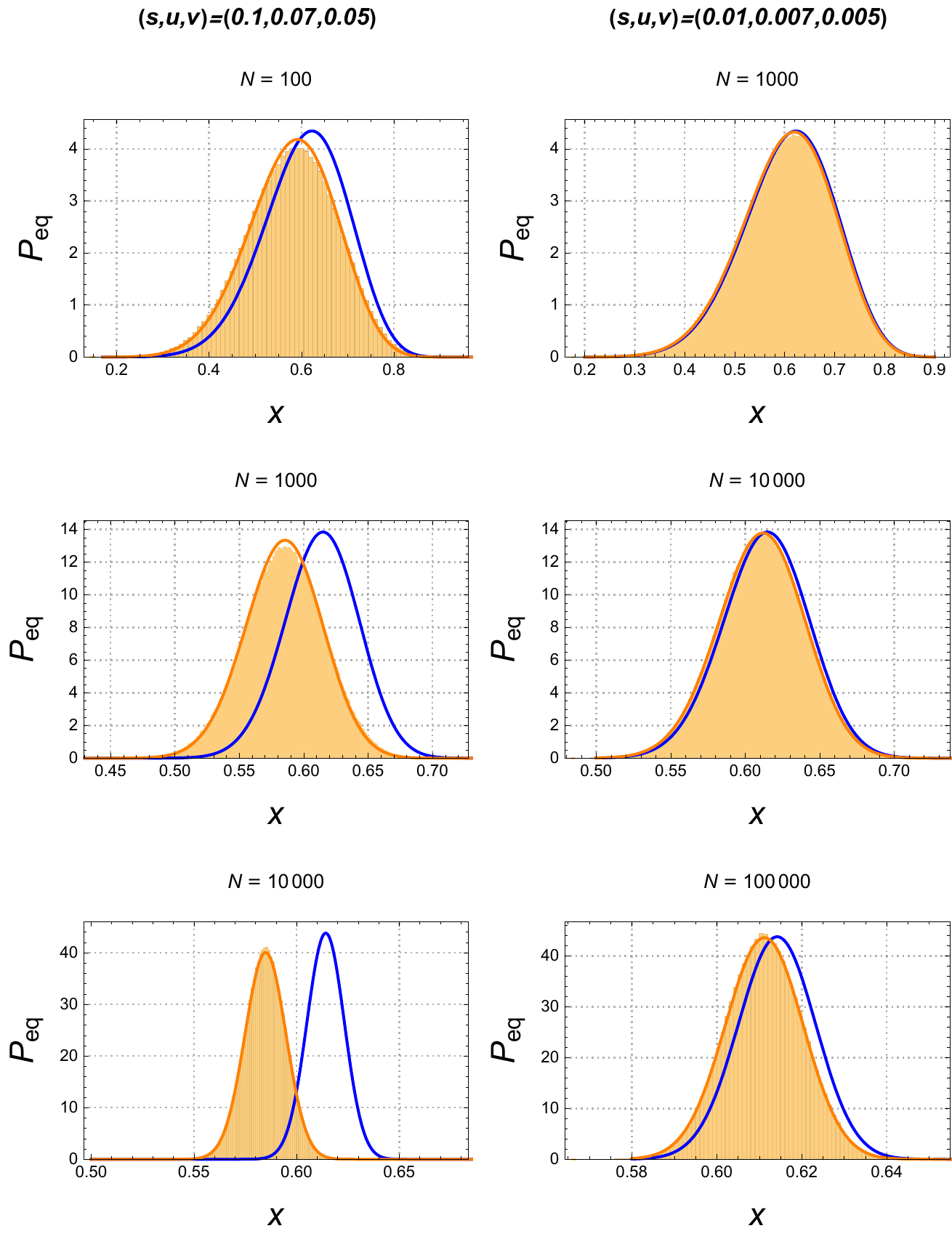}
    \caption{Stationary distributions as result of $10^6$ simulated generations of the WF process (Eq.~\eqref{WF}, yellow histograms) compared to analytical predictions from the WF diffusion (Eq.~\eqref{WFdiff}, blue line) and the diffusion by interpolation (Eq.~\eqref{newDiff}, orange line). The parameters $s,u,v$ as given in the subplots are kept constant when moving vertically downwards where population size $N$ increases. Deviations of the WF diffusion from the underlying discrete model become apparent for strong selection $N\gg 1/s^2$.}
\label{stationary_distrib}
\end{figure}

\begin{figure*}
    \centering
    \includegraphics[width=\textwidth]{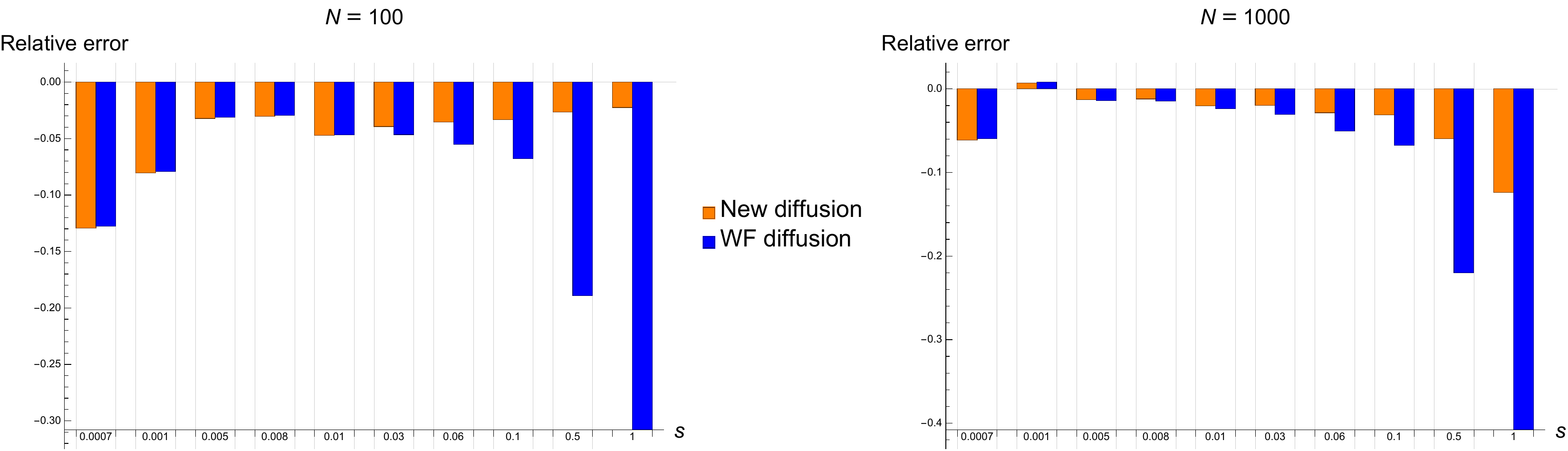}
    \caption{Relative errors in approximations of the establishment time of a rare mutant subject to reversions ($s,u \neq 0$). The errors of the two analytical approximations, WF diffusion (Eq.~\eqref{WFdiff}, in blue) and diffusive approximation by interpolation (Eq.~\eqref{newDiff}, in orange) are computed relative to results from simulations of the discrete WF process (Eq.~\eqref{WF}).}
    \label{fig:establishmentTimes}
\end{figure*}

In the limit of large population sizes, the standard diffusion approximation considers the WF model on the slow timescale $T=nN$, and assumes that selection and mutation are comparable in strength to genetic drift, i.e. $Ns$, $Nu$, and $Nv$ have finite limits $\sigma, \mu, \nu$ as $N\to\infty$. Under these conditions, it can be shown that the paths of the WF model on the slow timescale converge to those of the diffusion process
\begin{eqnarray}
    dX_T = [\sigma X_T(1-X_T) - \mu X_T + (1-\nu)X_T]dT + \sqrt{X_T(1-X_T)}dW_T
\end{eqnarray}
where $dW_T$ denotes a standard Wiener process and the Ito convention for stochastic differential equations is used. When brought back to the fast timescale $t=T/N$, this process becomes
\begin{eqnarray}
    dX_t = [sX_t(1-X_t) - uX_t + (1-v)X_t]dt  + \sqrt{\frac{X_t(1-X_t)}{N}}dW_t.
      \label{WFdiff}
\end{eqnarray}
Eq.~\eqref{WFdiff} above is the ``WF diffusion equation" found in textbooks. By construction, this approximation relies on the condition that all evolutionary parameters scale as $1/N$ when $N$ is taken to infinity, corresponding to the ``weak selection-weak mutation" regime of evolution \citep{Ewens_2004}. A rule of thumb due to Nei states that selection can be considered small if $Ns^2 \ll 1$ \cite{nei2005selectionism, Assaf_2011}.

There is, however, another way to approximate the WF model with a diffusion process. By the central limit theorem, Eq.~\eqref{WF} may be written as
\begin{equation}
X_{n+1} \sim p(X_{n}) + \sqrt{\frac{p(X)(1-p(X_n))}{N}}\nu
\end{equation}
where $\nu$ approaches a standard normal variable if $N\gg 1$. The latter equation is the Euler-Maruyama discretization of the Ito diffusion equation
\begin{equation}
  dX_t = [p(X_t) - X_t]dt + \sqrt{\frac{p(X_t)(1-p(X_t))}{N}}dW_t
  \label{diff2}
\end{equation}
at times $t = 0, 1, \cdots$. If the fraction $X_t$ changes only slightly during each time increment (i.e. if $s,u,v \ll 1\ll N$), then Eq.~\eqref{diff2} is an accurate \emph{interpolation} of the original process, Eq.~\eqref{WF}. We simplify Eq.~\eqref{diff2} further as
\begin{equation}
    dX_t = [p(X_t) - X_t]dt + \sqrt{\frac{X_t(1-X_t)}{N}}dW_t.
    \label{newDiff}
\end{equation}
Versions of Eq.~\eqref{newDiff} appear in e.g. \citep{Waxman_2010, Manhart_2012}, but, to our knowledge, have never been used to derive analytical results valid beyond the weak selection-weak mutation limit of evolutionary dynamics. A mathematical discussion of neutral diffusions at large mutation rates can be found in \cite{Norman_1975, ethier1977error}.

\section{Stationary distributions and establishment times}

\subsubsection*{Stationary distributions}
That the diffusive approximation in Eq.~\eqref{newDiff} has a broader domain of validity than the standard WF diffusion in Eq.~\eqref{WFdiff} can be seen by comparing their stationary distributions. We recall that, given a stochastic differential equation of the kind described above,
\begin{equation}
     dX_t = a(X_t) \, dt + \sqrt{b(X_t)} \, dW_t,
\end{equation}
a stationary distribution $P_{\text{eq}}$ exists as long as both boundaries, $x=0$ and $x=1$, are reflecting ($v\neq 0$, $u\neq0$) and can be calculated as \citep{van1992stochastic}
\begin{equation}
     P_{\text{eq}} (x)\propto \frac{\exp{\left[2\int ^x\frac{a(y)}{b(y)} \diff y\right]}}{ b(x)}.
\end{equation}
In Fig. \ref{stationary_distrib} we compare the analytical expressions from both diffusive approximations for $P_{\text{eq}}$ with simulations of the WF process with $p(x)$ from Eq.~\eqref{probability}. Results are shown for a range of parameters $s, u, v$, both within and outside the weak selection-weak mutation regime defined by $s,u,v = \mathcal{O}(1/N)$.

\begin{figure*}
    \begin{center}
    \includegraphics[width=\textwidth]{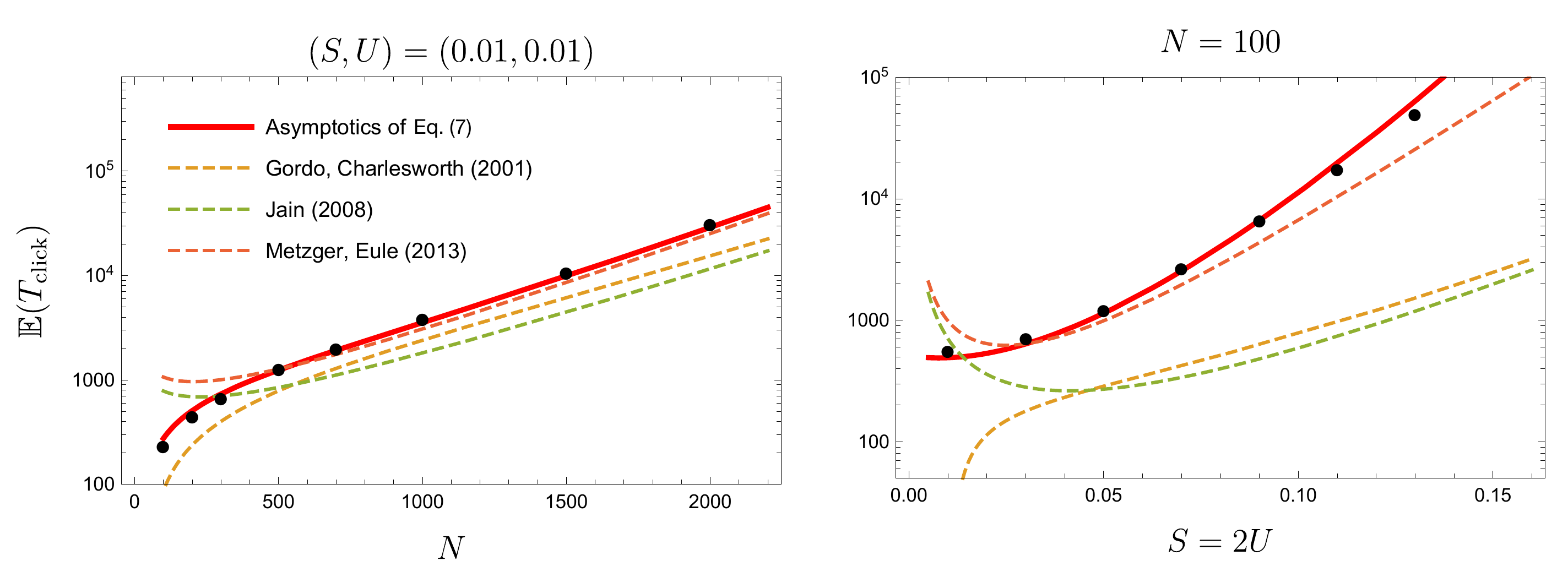}
    \caption{Accuracy of different analytical expressions for the mean time between clicks of Muller's ratchet in the slow click regime, as a function of $N$ for $S=U=0.01$ (left) and as a
    function of $S=2U$ for $N=100$ (right). Our
    result (Eq.~{\eqref{clickTime}}, red line), derived from the diffusion
    interpolation (Eq.~{\eqref{newDiff}}) through Laplace's method
    correctly approximates Wright-Fisher simulations (mean click time
    over $10^3$ replicates, represented as black dots) of the two-type reduced model of Muller's ratchet, including when selection and mutation are strong
    ($S=U\gg1/N$).}
    \label{FigRatchetRate}
    \end{center}
\end{figure*}

\subsubsection*{Establishment times with one-way mutations}
The classical fixation problem in population genetics considers the fate of a single selectively advantageous mutant ($s>0$) subject to finite population size fluctuations (genetic drift) only. Mutation rates are zero ($u=v=0$). Diffusive approximations can be used in that case to derive an analytical estimate for the probability of fixation at the boundaries $x=0$ or $x=1$ and corresponding expected times to fixation. For completeness, we report the results of the classical fixation problem in Appendix {\ref{sec:appendix}}, but note that the differences between predictions from the two diffusive schemes are only minor. 

Instead, we focus here on a modified fixation problem in which the advantageous mutant is additionally subjected to one-way mutations ($u \neq 0$, $v=0$). As in the classical setting, when a single mutant $X_0=1/N$ appears in the population, the probability of immediate extinction is high (due to genetic drift). Still, there is a finite probability of reproduction and establishment. Establishment means here reaching a metastable state $x_c>0$, defined by $p(x_c)=x_c$. The boundary $x=1$ is no longer absorbing due to reversions at a finite rate $u$ and the mutant will almost certainly go extinct in finite time, since $x=0$ represents the only absorbing boundary. The intermittent dynamics may still be of interest and can be characterised by the probability and time of establishment.

In Fig. \ref{fig:establishmentTimes} we compare the relative errors of mean establishment times $\overline{t}_{\textrm{est}}$ between WF simulations and predictions derived from the two diffusive approximations (Eq.~\eqref{WFdiff} and Eq.~\eqref{newDiff}). Details for the calculations are found in the Appendix. Here, again, the relative errors for the prediction of the WF diffusion increase outside the weak selection regime $s^2>1/N$, unlike the estimates derived from the diffusive approximation by interpolation (Eq.~\eqref{newDiff}). Details of the calculations can be found in the Appendix \ref{sec:appendix}.

\section{Rate of Muller's ratchet}
Haigh's model of Muller's ratchet \citep{Haigh_1978} considers a
population of fixed size $N$ where deleterious mutations accumulate
at a constant rate $U$, such that an individual
carrying $k$ mutations has reduced fitness $\left(1-S\right)^k$.
When $Ne^{-U/S}>1$, fixations are rare and clicks of the ratchet
are large deviations (rare fluctuations exponentially suppressed
in $N$) this is commonly referred to as the slow
click regime. Although significant progress was made in the last
decade \citep{Neher_2012}, Haigh's model is difficult to study
analytically, and a common approach is to approximate it by a two-type
WF model  \citep{Waxman_2010,Assaf_2011,Metzger_2013}. In this simpler model, one class corresponds to the fittest (unmutated) individuals, with unit fitness, and all other mutants are collectively assigned a reduced fitness $(1-s)$ with $s=(1-e^{-U})/(1-e^{-U/S})$; the effective
mutation rate away from the fittest class is then $u=1-e^{-U}$.
Assuming fecundity selection following \cite{Assaf_2011,Metzger_2013}, we arrive at the WF model in Eq.~{\eqref{WF}} with sampling
probability $p(x)=x(1-u)/(1-s+sx)$. In contrast to Eq.~{\eqref{probability}} the selection coefficient is redefined as a coefficient of negative selection, hence the normalisation term of mean fitness is adjusted accordingly.

We used classical expressions for hitting times of one-dimensional diffusions \cite{Ewens_2004} combined with Laplace's approximation of integrals to derive from Eq.~{\eqref{newDiff}} an analytical expression for the expected time between clicks of the ratchet; technical details are provided in the Appendix. On a
logarithmic scale, this expression reads
\begin{equation}
\log\mathbb{E}(T_{\textrm{click}})\underset{N\to\infty}{\sim}2N\log\left[\left(\frac{u}{s}\right)^u\left(\frac{1-u}{1-s}\right)^{1-u}\right],
    \label{clickTime}
\end{equation}
compared to the approximation that is found by \cite{Metzger_2013}
\begin{equation}
    \log\mathbb{E}(T_{\textrm{click}})\sim 2N[s-u+u\log (u/s)],
    \label{metzgereule}
\end{equation}
and corresponds to the expected hitting time of the WF diffusion (Eq.~{\eqref{WFdiff}} instead of Eq.~{\eqref{newDiff}}). Fig. \ref{FigRatchetRate} compares the accuracy of our result with earlier heuristic diffusion approximations \citep{GORDO_2001,Jain_2008} as well as more recent WKB expansions of Moran models \citep{Metzger_2013}. The near-perfect agreement in the large $N$ limit---including when $S,U\gg 1/N$---confirms that diffusion methods are more general than scaling limits and can be successfully applied outside the weak selection-weak mutation regime of evolution.

\section{Conclusion}
Because it does not depend on the mode (fecundity or viability) of selection, on generations being overlapping or non-overlapping, etc., the classical WF diffusion equation \eqref{WFdiff} is sometimes described as ``universal''. What this means is simply that, in the special regime where $s,u,v$ are all very small relative to the finite size fluctuations ($s, u, v = \mathcal{O}(1/N)$), these differences are immaterial for the long-term dynamics of the population. But that is no longer true in regimes where selection and/or mutation are much stronger than genetic drift. In these regimes, diffusion theory does not break down; instead, other, non-universal diffusion approximations of the WF model can be used to obtain accurate results for fixation times and other quantities of interest. 

Importantly, diffusion equations such as Eq.~\eqref{newDiff} can be used to estimate the timescales of large deviations, as observed e.g. in the slow clicking regime of Muller's ratchet. This conclusion is in contrast with previously expressed concerns regarding the use of diffusion theory in the context of large deviations \citep{doering2005extinction,ovaskainen2010stochastic,hanggi1984bistable,kessler2007extinction}: ``The fact that the extinction of the fittest class is due to such a rare, large fluctuation and not the cause of a typical fluctuation prohibits simple diffusive treatments of the ratchet'' \citep{Metzger_2013}. Our results show, instead, that the analytical approximation of Muller's ratchet within a diffusive framework improves previous results in the strong selection regime $s>1/\sqrt{N}$. We expect that applications of diffusive approximations beyond the ones here presented might allow further progress in the analysis of evolutionary processes that are well captured by the discrete-time Wright-Fisher model, but not by its textbook diffusion approximation.

\section*{Acknowledgments}
We thank Maseim Kenmoe for help with Laplace integrals, Aleksander Klimek and Anton Zadorin for comments on the manuscript, and Sophie P\'enisson for pointing us to the mathematical references \cite{Norman_1975, ethier1977error}. Funding for this work was provided by the Alexander von Humboldt Foundation in the framework of the Sofja Kovalevskaja Award endowed by the German Federal Ministry of Education and Research.

\appendix

\section{Appendix}\label{sec:appendix}

\subsection{Derivation of Muller's ratchet rate formula}
Eq.~\eqref{newDiff} from the main text describes the dynamics of the stochastic variable $X_t$ in terms of a SDE. Moving to an alternative dynamical description in terms of the probability density function $\rho (x,t)$ of the stochastic variable $x$ at time $t$, the corresponding drift-diffusion equation (also Fokker-Planck equation or Kolmogorov forward equation) takes the following form
\begin{equation}
    \label{generalDriftDiffusion}
    \frac{\partial \rho(x,t)}{\partial  t} = - \frac{\partial  }{\partial  x} \left[ a(x) \, \rho(x,t) \right] + \frac{1}{2} \frac{\partial^2 }{\partial x^2} \left[ b(x)  \, \rho(x,t) \right] ,
\end{equation}
where in our setting $a(x) = p(x) -x $ and $b(x)  = x(1-x)/N$. \\
We are interested in the expected time for the stochastic variable $x$ to hit the boundary at $x=0$, given an initial position $x=x_0$ (to be specified below).
In the scope of diffusion theory this is referred to as the mean first passage time $\overline{t}(x_0)$, which is described by a second order differential equation \citep{van1992stochastic}
\begin{equation}
    \label{MeanFirstPassageTimes}
     a(x_0) \frac{\diff \overline{t}(x_0)}{\diff x_0} + \frac{1}{2} b(x_0) \frac{\diff^2 \overline{t} (x_0)}{\diff x_0^2} = -1 .
\end{equation}
In the setting of Muller's ratchet, the diffusion of Eq.~\eqref{generalDriftDiffusion} is bounded by definition to the interval $[0,1]$ (since $x$ describes a relative frequency). Solving Eq.~\eqref{MeanFirstPassageTimes}, subject to the appropriate boundary conditions for reflection at $x=1$ and absorption at $x=0$, namely
\begin{equation*}
   \frac{\diff \overline{t}(x)}{\diff x} \rvert_{x=1}  = 0 \qquad \text{and} \qquad
      \overline{t}(0) =0 ,
\end{equation*}
the expected click time of Muller's ratchet can be expressed in terms of the following equation
\begin{equation}
\begin{split}
     \overline{t}(x_0) &= \int_0^1 \frac{2}{b(x)}  \exp{\left(\Psi(x) \right)}\,  \times \\
     &\int_0^{\min [x,x_0]} \exp{\left(-\Psi(y)\right)} \diff y \diff x  ,
\end{split}
\label{MRIntegralForm}
\end{equation}
where $\Psi(x) = \int^x \frac{2a(z)}{b(z)} \diff z $. Eq.~\eqref{MRIntegralForm} cannot be solved analytically, but instead  the proportionality of the argument in the exponential $\Psi(x) \propto N $ to the population parameter $N$, renders the integral equation suitable for a Laplace approximation. We therefore expand $\Psi(x)$ in a Taylor series around a critical point $x_c$ and truncate the expansion by assuming $(x-x_c) = \mathcal{O}\left(\frac{1}{\sqrt{N}}\right)$, so that
\begin{equation}
\begin{split}
     \Psi(x)&= \Psi(x_c) + \Psi'(x_c)\left(x-x_c\right)\\&+ \frac{1}{2} \Psi''(x_c) \left(x-x_c\right)^2 + \mathcal{O}\left(\frac{1}{\sqrt{N}}\right) ,
\end{split}
\label{PsiExpansion}
\end{equation}
since $N (x-x_c)^3 = \mathcal{O}\left(\frac{1}{\sqrt{N}}\right)$.
Hence, the exponential integral is approximately reduced to a Gaussian integral on a bounded interval, which can be evaluated in terms of error functions. Generally speaking,
\begin{widetext}
\begin{equation}
\label{LaplaceApprox}
\begin{gathered}
\int_a^b \exp\left(\Phi(x) \right) \approx \int_a^b  \exp\left[\Phi(x_c) + \Phi'(x_c)\,(x-x_c) + \frac{1}{2} \Phi''(x_c)\, (x-x_c)^2\right] \diff x \\
  =C \times \, \int_a^b \exp\left[\left( \sqrt{\frac{\Phi''(x_c)}{2}}(x-x_c) + \frac{\Phi'(x_c)}{\sqrt{2 \Phi''(x_c)}}\right)^2\right] \diff x \\
 = C\times \sqrt{\frac{2}{\Phi''(x_c)}} \iu \int_{\iu \frac{(a-x_c) \Phi''(x_c) + \Phi'(x_c)}{\sqrt{2 \Phi''(x_c)}}}^{\iu \frac{(b-x_c) \Phi''(x_c) + \Phi'(x_c)}{\sqrt{2 \Phi''(x_c)}}} \exp(-\chi^2) \, \diff \chi \\
 = C \sqrt{\frac{\pi}{2 \Phi''(x_c)}}\left( \erfi \left[\frac{(b-x_c) \Phi''(x_c) + \Phi'(x_c)}{\sqrt{2 \Phi''(x_c)}} \right] - \erfi \left[ \frac{(a-x_c) \Phi''(x_c) + \Phi'(x_c)}{\sqrt{2 \Phi''(x_c)}} \right] \right) ,
\end{gathered}
\end{equation}
\end{widetext}
   where $C =\exp\left[\Phi(x_c) - \frac{\Phi'(x_c)^2}{2 \Phi''(x_c)}\right]$ and $\erfi(x)$ refers to the imaginary error function.\\
Considering in Eq.~\eqref{MRIntegralForm} an additional expansion of the non-exponential function around the critical point $g(x) = \frac{2}{b(x)} = g(x_c) + \mathcal{O}\left(\frac{1}{\sqrt{N}}\right)$, as well as an approximation of the limit $\min[x,x_0] \approx \min[x_c,x_0] $ we notice that the previously nested integrals decouple and we can deduce the full approximation by applying the Laplace approximation (Eq.~\eqref{LaplaceApprox}) to both integrals separately
\begin{widetext}
\begin{equation}
    \begin{gathered}
     \overline{t}(x_0) = \int_0^1 \frac{2}{b(x)}  \exp{\left(\Psi(x) \right)} \int_0^{\min [x,x_0]} \exp{\left(-\Psi(y)\right)} \diff y \diff x \\
     \approx g(x_c) \frac{\pi}{2\sqrt{-\Psi''(y_c)\Psi''(x_c)}} \exp\left[\Psi(x_c) - \frac{\Psi'(x_c)^2}{2 \Psi''(x_c)} - \Psi(y_c) +\frac{\Psi'(y_c)^2}{2 \Psi''(y_c)}\right] \times \\
     \left( \erfi \left[ \frac{-(\min[x_c,x_0]-y_c) \Psi''(y_c) - \Psi'(y_c)}{\sqrt{- 2 \Psi''(y_c)}} \right] - \erfi \left[ \frac{-(0-y_c) \Psi''(y_c) - \Psi'(y_c)}{\sqrt{-2 \Psi''(y_c)}} \right] \right) \times \\
     \left( \erfi \left[ \frac{(1-x_c) \Psi''(x_c) + \Psi'(x_c)}{\sqrt{2 \Psi''(x_c)}} \right] - \erfi \left[ \frac{(0-x_c) \Psi''(x_c) + \Psi'(x_c)}{\sqrt{2 \Phi''(x_c)}} \right] \right).
    \end{gathered}
\end{equation}
\end{widetext}
For the ratchet click time in the slow-clicking regime, we assume the initial position to be precisely at the deterministic equilibrium, i.e. where $a(x_0) = x_0 \left(\frac{1-u}{1-s +s x_0}-1\right) \overset{!}{=} 0$, which means the initial point $x_0$ and the critical point $x_c$ of the potential $\Psi(x)$ coincide, $x_0 = x_c = 1-u/s$. It is from this, that the condition for the slow clicking regime can be deduced, namely $u<s$, which implies the existence of a deterministic equilibrium. \\We arrive finally at the following expression for the expected click time in terms of our evolutionary parameters $N,s,u$
\begin{widetext}
\begin{eqnarray}
    \label{FullLaplaceSolution}
    \begin{gathered}
     \mathbb{E}(T_{\textrm{click}}) \approx \frac{\pi  s^2\exp \left(2N \left(\frac{(s-u)^2}{-2s^2+4 s u-2u}+ u \log \left(\frac{u}{s}\right)+(1-u) \log \frac{1-u}{1-s}\right)\right)}{2 u \sqrt{\frac{s^4}{(s-1)^2 u}-\frac{s^2}{u-1}} (s-u)} \times \\ \left(\erf\left(\frac{(s-u) \sqrt{-\frac{N (s-1)^2}{s^2-2 s u+u}}}{s-1}\right)-\erf\left(\frac{(2 s-1) (s-u)^2 \sqrt{-\frac{N (s-1)^2}{s^2-2 s u+u}}}{(s-1)^2 s}\right)\right) \times \\ \left(\erfi\left(\frac{N \left(1-\frac{u}{s}\right)}{\sqrt{\frac{N (u-1) u}{s^2}}}\right)+\erfi\left(\frac{N u}{s \sqrt{\frac{N (u-1) u}{s^2}}}\right)\right) .
    \end{gathered}
\end{eqnarray}
\end{widetext}
On a logarithmic scale, Eq.~\eqref{FullLaplaceSolution} yields in the limit of large populations
\begin{equation}
\label{clickTime2}
  \log\mathbb{E}(T_{\textrm{click}})\underset{N\to\infty}{\sim}2N\log\left[\left(\frac{u}{s}\right)^u\left(\frac{1-u}{1-s}\right)^{1-u}\right]  ,
\end{equation}
in accordance with Eq.~\eqref{clickTime} from the main text.

\begin{figure*}
    \centering
    \includegraphics[width=\textwidth]{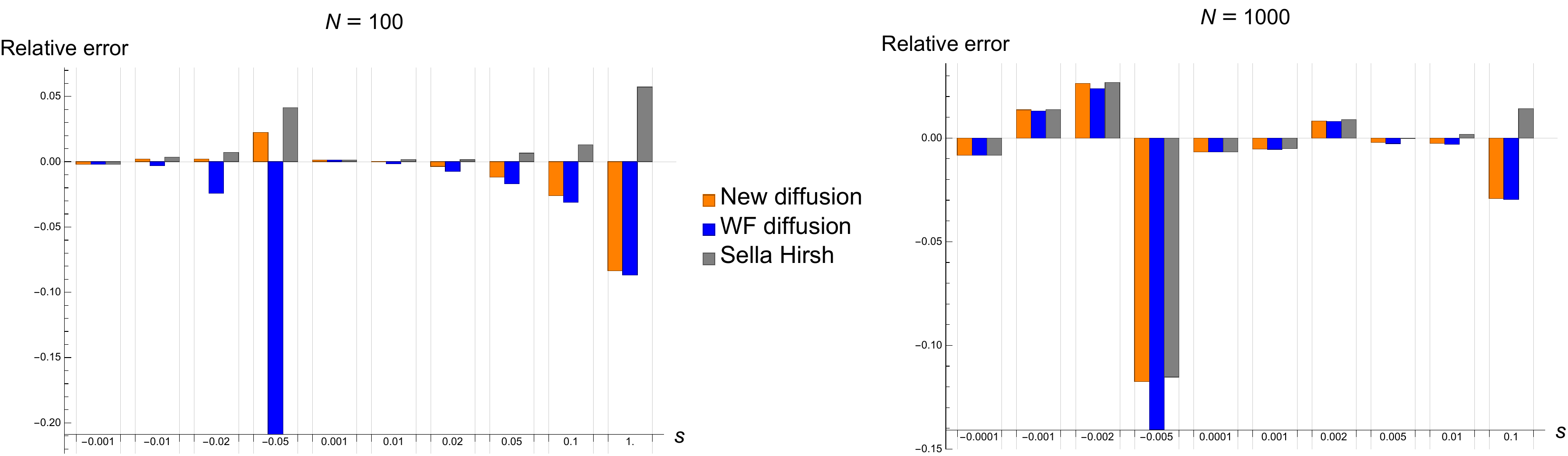}
    \caption{Errors in predicted fixation probabilities to the classical fixation problem over varying selection coefficients $s$ and population sizes $N=100, 1000$. Approximations by \cite{Sella_2005} (Eq.~\eqref{eq:fixprobSH}, in grey), \cite{kimura1962probability} in the WF diffusion scheme (Eq.~\eqref{eq:fixprobKimura}, in blue), and from the new  diffusion scheme (Eq.~\eqref{eq:fixprobNewDiff}, in orange) are compared to relative number of fixations in $10^8$ ($N=100$) and $10^7$ ($N=1000$) WF simulations.}
    \label{fig:FixProbab}
\end{figure*}

An analogous procedure, in which we replace $a(x)= sx(1-x)-ux$ from the WF diffusion, Eq.~\ref{WFdiff}, can be used to derive an estimate for Muller's ratchet rate from the WF diffusion
\begin{widetext}
\begin{eqnarray}
\begin{gathered}
\mathbb{E}(T_{\textrm{click}}) \approx \frac{\pi s}{2(s-u) u}  \exp{\frac{- N (s-u)^2}{u}} \times \exp{\left(2 N [s-u +u \log\Big[\frac{u}{s}\Big]] \right)} \times \\
 \quad \left[ \mathrm{erf} \left(\sqrt{\frac{-N}{u}} \frac{(s - u)^2}{s} \right) - \mathrm{erf} \left(\sqrt{\frac{-N}{u}} (s-u) \right) \right] \times
 \quad \left[ \mathrm{erfi} \left( \sqrt{-Nu} \left( \frac{s}{u}- 1 \right) \right) + \mathrm{erfi} \left(\sqrt{-Nu} \right) \right] \,.
 \end{gathered}
\end{eqnarray}
\end{widetext}
For $\sigma < 1-\frac{u}{s}$, where $\sigma^2= \frac{u}{2Ns^2}$ denotes the variance of the Gaussian, the approximation can be reduced to
 \begin{equation}
\mathbb{E}(T_{\textrm{click}}) \approx \sqrt{\frac{\pi}{Nu}} \frac{s}{(s-u)^2} \exp{ \big(2N(s-u+\log\Big[\frac{u}{s}\Big]) \big)} ,
\label{MRTimeOld}
    \end{equation}
which is the same equation as derived by \cite{Metzger_2013} through WKB approximations of the associated Moran process (after a rescaling of time and population size $t' \rightarrow t/2N$
and $N' \rightarrow 2N$). The result in Eq.~\eqref{MRTimeOld} coincides on a logarithmic scale with Eq.~\eqref{metzgereule}, as described in the main text.

\subsection{Fixation Probability and Fixation Times}

Additionally, we here report the comparison of the estimated fixation probabilities and fixation times for the Wright-Fisher model without mutations $u=v=0$. In Fig. \ref{fig:FixProbab} we compare both diffusive approximations, Eq.~\eqref{WFdiff} and Eq.~\eqref{newDiff}, as well as the formula reported in \cite{Sella_2005}. We notice that the corrections in the strong selection regime are minor. Analytical formulae are compared to mean fixation probabilities from a total of $10^8$ (for $N=100$) and $10^7$ (for $N=1000$) Wright-Fisher simulations (Eq.~\eqref{WF}).
The estimate from the WF diffusion, Eq.~\eqref{WFdiff} is the classical formula for the fixation problem as presented by Crow and Kimura \cite{kimura1962probability}
\begin{equation}
    \pi_{\textrm{CK}}=\frac{1-e^{-2s}}{1-e^{-2Ns}}.
      \label{eq:fixprobKimura}
\end{equation}

More recently, it was shown in \cite{Sella_2005} that
\begin{equation}
    \pi_{\textrm{SH}}=\frac{1-(1+s)^{-2}}{1-(1+s)^{-2N}}
    \label{eq:fixprobSH}
\end{equation}
is slightly more accurate that the
Crow-Kimura formula, among other good properties.

Making use of the diffusive scheme by interpolation, Eq.~\eqref{newDiff}, given an initial point $x=x_0$, the
probability of fixation at $x=1$ is approximated by
\begin{equation}
    \pi(x_0)=\frac{(1+s)^{2N}(1+x_0 s)^{-2N}((1+x_0 s)^{2N}-1-x_0 s)}{(1+s)^{2N}-1-s},
    \label{eq:fixprobNewDiff}
\end{equation}
which reduces to the classical
fixation problem for $x_0=1/N$.

\begin{figure*}
    \centering
    \includegraphics[width=\textwidth]{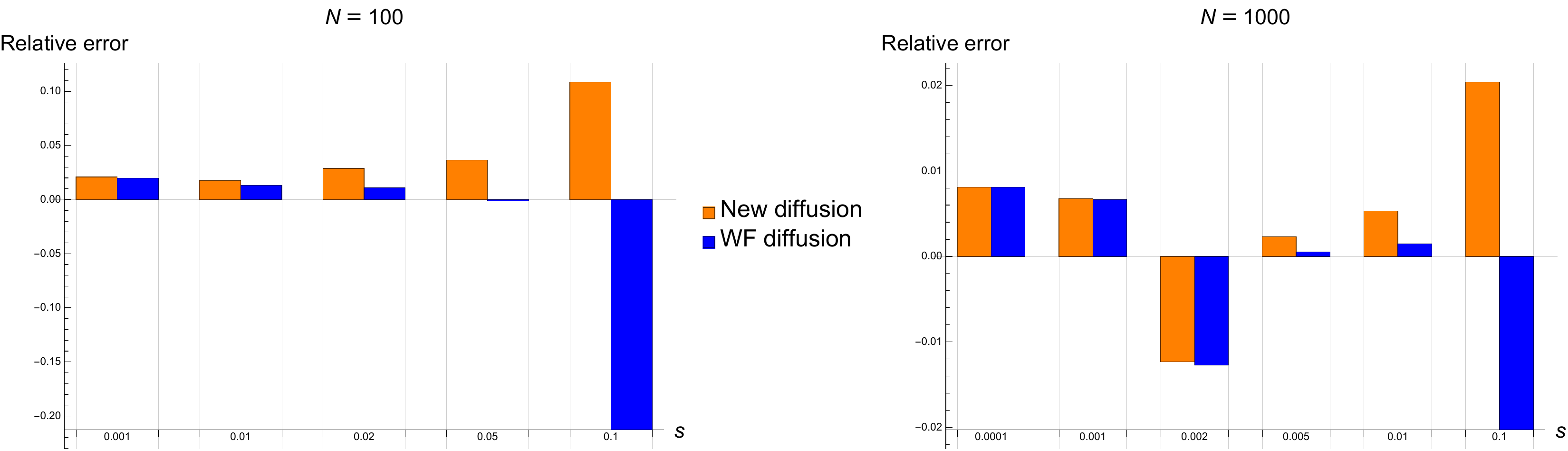}
    \caption{Relative errors in approximations of fixation times for varying selection coefficients $s$ and population sizes $N=100, 1000$ as calculated from Eq.~\eqref{eq:fixTime}, for the WF diffusion (in blue) and the new diffusion (in orange). Errors are computed relative to simumlated mean fixation times over $10^8$ ($N=100$) and $10^7$ ($N=1000$) runs of of WF simulations.}
    \label{fig:fixTimes}
\end{figure*}

Using Eq.~\eqref{MeanFirstPassageTimes} with the appropriate boundary conditions $\overline{t}(0)=\overline{t}(1)=0$
(two absorbing boundaries), the mean first hitting time at $x=1$, conditional on fixation at $x=1$, with initial condition $x=x_0$ can be computed through the following integral \citep{van1992stochastic}
\begin{equation}
    \begin{split}
         \overline{t}(x_0)= \frac{\pi^0(x_0)}{\pi^1 (x_0)} \int_0^{x_0} \frac{2  \pi^1 (x)}{ b(x) \Psi(x)} \int_0^x \Psi(y)\, \diff y\, \diff x \\
     + \int_{x_0}^1 \frac{2 \pi^1(x)}{b(x) \Psi(x)}\, \int_x^1 \Psi(y) \, \diff y\,\diff x,
    \end{split}
    \label{eq:fixTime}
\end{equation}
with $\Phi(x) = \int^x \frac{-2a(z)}{b(z)} \diff z$. Herein, $\pi^1(x)$ and $\pi^0(x)$ denote the fixation probabilities at $x=1$
and $x=0$, respectively, as a function of an initial point $x_0$. The relative errors in estimates deriving from the differing $a(x)$ in the WF diffusion (Eq.~\eqref{WFdiff}) and the diffusive approximation by interpolation (Eq.~\eqref{newDiff}) are compared in Fig. \ref{fig:fixTimes} relative to mean times of fixation from WF simulations for varying parameters $s, N$.

\subsection{Mean establishment time with reversions}
As presented in Fig. \ref{fig:establishmentTimes} of the main text, analytical estimates between the two diffusive approximations for the Wright-Fisher model with selection and one-way mutations $v=0$ are compared. These formulae can be derived equivalently to the above described fixation times. The probability of establishment, given an inital point $x_0$, is given by
\begin{equation}
    \pi^{x_c} (x_0)=\frac{\int_0^{x_0}  \Phi(y) \diff y}{\int_0^{x_c}  \Phi(y) \diff y},
\end{equation}
where as before $\Phi(x) = \int^x \frac{-2a(z)}{b(z)} \diff z$ and $x_c$ denotes the metastable state defined by $p(x_c)=x_c$.
\\

From this the expected time to establishment , assuming the initial frequency $x_0=\frac{1}{N}$, is computed by
\begin{equation}
\begin{split}
     \overline{t}_{\text{est}}= \frac{2 \left(1-\pi^{x_c}\left(\frac{1}{N}\right)\right)}{\pi^{x_c} \left(\frac{1}{N}\right)} \int_0^{\frac{1}{N}} \int_0^x  \frac{\Phi(y) \pi^{x_c}(x)}{b(x) \Phi(x)} \diff y \diff x \\+
     \int_{\frac{1}{N}}^{x_c}    \int_{x}^{x_c} \frac{2 \Phi(y) \pi^{x_c}(x)}{b(x) \Phi(x)}  \diff y \diff x \,.
\end{split}
\end{equation}

\subsection{WF simulations}
Stochastic simulations were run on Wolfram Mathematica according to the dynamics from Eq.~{\eqref{WF}}, while varying the evolutionary
parameters $N,s,u, v$ as given in the main text. Due to limited computation times only a constrained set of parameters are tested that do not exceed computational feasibility.
\vspace{2cm}

\bibliographystyle{apsrev4-1}
\bibliography{biblio_diffusiontheory.bib}

\end{document}